\documentclass[aps,prl,10pt,twocolumn,amsmath,showpacs,showkeys,superscriptaddress]{revtex4-1}
\usepackage{graphicx}
\usepackage{color}
\usepackage{bm}
\usepackage{dsfont}
\usepackage{braket}

\begin{document}

$\newcommand{\E}{\mathrm{E}}$

$\newcommand{\Var}{\mathrm{Var}}$

$\newcommand{\Cov}{\mathrm{Cov}}$


\title{Stabilizing nuclear spins around semiconductor electrons via the interplay of\\ optical coherent population trapping and dynamic nuclear polarization}


\author{A.~R.~Onur}\thanks{These authors contributed equally to this work.}
\author{J.~P.~de~Jong}\thanks{These authors contributed equally to this work.}
\author{D.~O'Shea}
\affiliation{Zernike Institute for Advanced Materials, University of Groningen, 9747AG Groningen, The Netherlands}
\author{D.~Reuter}\thanks{Now at Department of Physics, University Paderborn, Warburger Stra{\ss}e 100, 30098 Paderborn, Germany.}
\author{A.~D.~Wieck}
\affiliation{Angewandte Festk\"{o}rperphysik, Ruhr-Universit\"{a}t
Bochum, D-44780 Bochum, Germany}
\author{C.~H.~van~der~Wal}
\affiliation{Zernike Institute for Advanced Materials, University of Groningen, 9747AG Groningen, The Netherlands}

\date{\today}

\begin{abstract}
We experimentally demonstrate how coherent population trapping (CPT) for donor-bound electron spins in GaAs results in autonomous feedback that prepares stabilized states for the spin polarization of nuclei around the electrons.
CPT was realized by excitation with two lasers to a bound-exciton state. Transmission studies of the spectral CPT feature on an ensemble of electrons directly reveal the statistical distribution of prepared nuclear spin states.
Tuning the laser driving from blue to red detuned drives a transition from one to two stable states.
Our results have importance for ongoing research on schemes for dynamic nuclear spin polarization, the central spin problem and control of spin coherence.
\end{abstract}


\maketitle
Following the emergence of electron spins in quantum dots and solid state defects as candidates for spin qubits it has become a major goal to realize control over the nuclear spins in such nanostructures.
In many experimental settings, interaction with disordered nuclear spins in the crystal environment is detrimental to the coherent evolution of carefully prepared electron spin states \cite{dyakonov1973, khaetskii2002, merkulov2002}. Preparation of nuclear spins in a state that has reduced spin fluctuations with respect to the thermal equilibrium state will help to overcome this problem \cite{coish2004}. Proposals to achieve this goal have been put forward for electron spin resonance (ESR) on one- or two-electron quantum dots \cite{danon2008,klauser2006}, and for optical preparation techniques that either rely on a quantum measurement technique \cite{giedke2006,stepanenko2006} or a stochastic approach \cite{issler2010,korenev2011,shi2013}. Experimental advances have been made with ESR and optical techniques on single quantum dots \cite{reilly2008,vink2009,latta2009,xu2009,bluhm2010,ladd2010,hansom2014} and nitrogen-vacancy centers \cite{togan2011}, and on quantum dot ensembles \cite{greilich2007,carter2009}.

Several of these works \cite{stepanenko2006,xu2009,issler2010,korenev2011,togan2011,shi2013,hansom2014} make use of the optical response of the electronic system near the coherent-population-trapping resonance (CPT, explained below) because it is highly sensitive to perturbations from nuclear spins. Notably, these experiments so far have focussed on quantum dots where, due to the particular anisotropic confinement, hyperfine coupling with a hole-spin in the excited state is reported to dominate \cite{xu2009}. In recent work \cite{onur2014} we discussed how the interplay between electron-nuclear spin interaction and CPT influences the stochastics of the nuclear spin bath for a class of systems where hyperfine interaction with the ground-state electron spin dominates.

Here we report experiments on this latter class of systems. We demonstrate an all-optical technique that stabilizes the nuclear spin bath around localized donor electrons in GaAs into a non-thermal state under conditions of two-laser optical pumping. We show that the nuclear spin system is directed either towards a single stable state or (probabilistically) towards one of two stable states, depending on laser detuning from the excited state. Our results show how feedback control arises from the interplay between CPT and dynamic nuclear spin polarization (DNP), and confirm that the electron-spin hyperfine interaction dominates for our system (despite the strong similarity with the negatively charged quantum dot). Our results indicate that this interplay can be used to create stable states of nuclear polarization with reduced fluctuations.

We perform measurements on the nuclear spin dynamics in a 10-$\rm{\mu m}$ thick MBE-grown film of GaAs doped with Si donors at a concentration of $\sim$3$\times$$10^{13}$~cm$^{-3}$, which is well below the metal-insulator transition (at $\sim$$10^{16}$~cm$^{-3}$). The wafer is cleaved in 2-by-2 mm$^{2}$ parts along the $\langle 1 1 0 \rangle$ crystal axes. The film is removed from a GaAs substrate by wet etching an AlAs buffer layer in HF. The film is then transferred to a sapphire substrate which allows us to do transmission measurements in a cryogenic microscope \cite{sladkov2011}. Measurements are performed at a temperature of $T=4.2$~K and magnetic field of $B_{ext}=5.9$~T. The sample is mounted such that the magnetic field direction is along the $\langle 1 1 0 \rangle$ axis. Light from tunable continuous-wave lasers (Coherent MBR-110) is delivered to the sample by a polarization-maintaining fiber and passes through the sample along the $\langle 1 0 0 \rangle$ axis. Transmitted light is collected in a multimode fiber and detected by an avalanche photodiode outside the cryostat. For getting reproducible data it was essential to stabilize laser powers within 1$\%$ and laser frequency drift within 10~MHz.

\begin{figure}
\centering
\includegraphics[width=7.5cm]{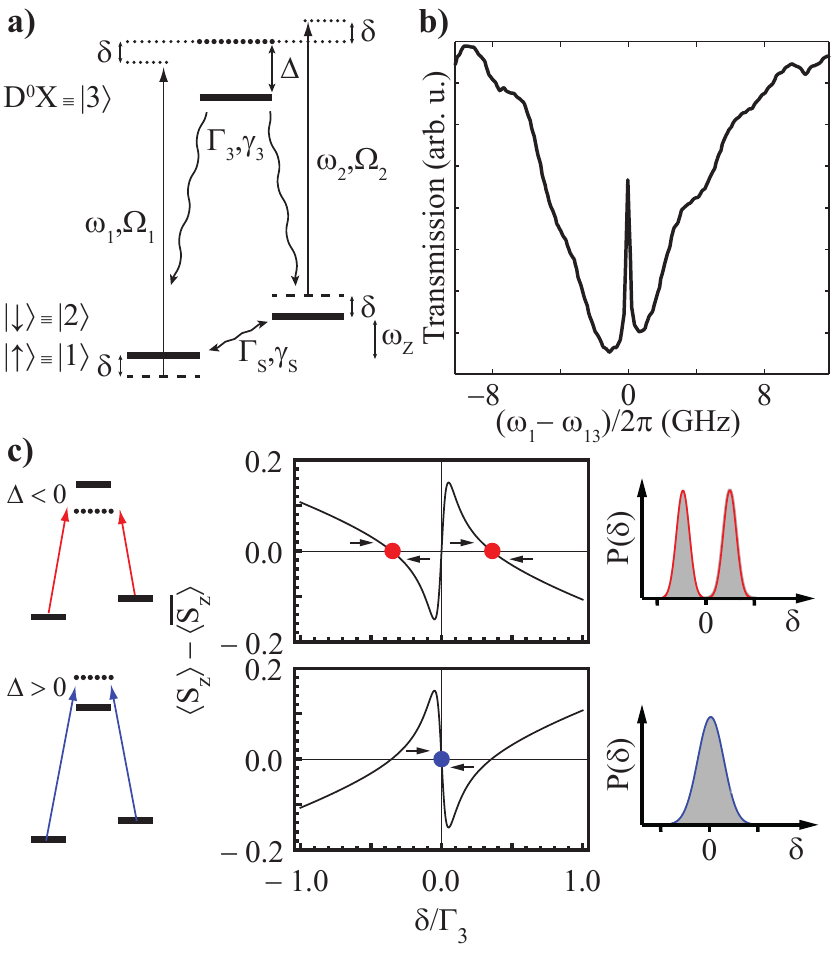}
\caption{Energy levels and feedback control scheme. (a) Thick black lines are the spin states $\ket{1}$, $\ket{2}$ and optically excited state $\ket{3}$. $\Gamma_{s}$, $\gamma_{s}$ and $\Gamma_{3}$ ,$\gamma_{3}$ are spin and excited state decay and dephasing rates, respectively. Two lasers at frequencies $\omega_{1}$, $\omega_{2}$ couple to the system with Rabi strengths $\Omega_{1}$ and $\Omega_{2}$, excited state detuning $\Delta$, and Overhauser shift $\delta$ (see main text). The energy splittings $\omega_{13}$, $\omega_{23}$ and $\omega_{Z}$ are fixed at the values for $\delta=0$ ($\hbar$ omitted for brevity). (b) Measured CPT signature in the $n$-GaAs sample (here for $\omega_{2}=\omega_{23}$ and $\delta=0$). (c) Left panel: two distinct control regimes for nuclear spin control. Middle panel: optically-induced electron spin polarization ($\braket{S_{z}}-\braket{\overline{S}_{z}}$) as a function of Overhauser shift $\delta$, with lasers fixed at $\omega_{1}=\omega_{13}+\Delta$ and $\omega_{2}=\omega_{23}+\Delta$, displays two stable states of the nuclear spin bath for $\Delta<0$ (red dots) and one stable state for $\Delta>0$ (blue dot). Right panel: expected steady state Overhauser shift distributions. Calculations with parameters $\gamma_{3}=10$, $\Gamma_{s}=10^{-4}$, $\gamma_{s}=10^{-3}$, $\Omega_{1}=\Omega_{2}=0.5$, $\Delta=\pm 1$ normalized to $\Gamma_{3}\equiv1$ \cite{onur2014}.}
\label{fig:lambda}
\end{figure}

The optical transitions that we address are from the donor-bound electron spin states ($\ket{\uparrow}$, $\ket{\downarrow}$) to a level of the bound trion ($\ket{D^{0}X}$), that consists of two electrons and one hole bound at the silicon donor. These three states form a $\Lambda$-type energy level configuration, further defined in Fig.~\ref{fig:lambda}(a). The magnetic field is applied perpendicular to the light propagation direction (Voigt geometry) such that the optical transitions have polarization selection rules discriminating between horizontally ($\sigma_{+,-}$, coupling to $\ket{\uparrow}-\ket{D^{0}X}$) and vertically ($\pi$, coupling to $\ket{\downarrow}-\ket{D^{0}X}$) polarized light.

We first demonstrate CPT for our system. CPT is a narrow resonance in two-laser driving as in Fig.~\ref{fig:lambda}(a) \cite{sladkov2010} where the system gets trapped in a dark state (for ideal spin coherence $\ket{\Psi} \propto \Omega_{2}\ket{\uparrow} - \Omega_{1}\ket{\downarrow}$). In transmission this appears as a narrow window of increased transparency within the broader absorption dip when one laser is scanning while the other is fixed (Fig.~\ref{fig:lambda}(b)). Its position signals two-photon resonance, and occurs where $\omega_{1}-\omega_{2}$ equals the the electron spin splitting. The lineshape of the CPT resonance can reveal information about the electron spin states, which can be obtained by fitting the curve in Fig.~\ref{fig:lambda}(b) to the Lindblad equation for the $\Lambda$-system \cite{fleischauer2005}. Our $n$-GaAs samples yield an inhomogeneous dephasing time $T_{2}^{*}\approx3$~ns \cite{sladkov2010}. However, the homogeneous dephasing time $T_{2}$ has been estimated to be at least $7~\rm{\mu s}$ \cite{clark2009} with a spin-echo technique. The discrepancy between $T_{2}$ and $T_{2}^{*}$ is largely due to dephasing caused by $\sim$$10^{5}$ disordered nuclear spins per electron.

Due to the Fermi contact hyperfine interaction, a non-zero nuclear spin polarization exerts an effective magnetic (Overhauser) field $B_{n}$ on the electron spin and causes a shift of the electron spin levels, denoted by $\delta$ in Fig.~\ref{fig:lambda}(a). The value of $\delta=p\delta_{\text{max}}$ is proportional to the nuclear spin polarization $p\in[-1,1]$, where $\delta_{\text{max}}$ is the maximum shift set by the hyperfine interaction strength. For the donor electron in GaAs $\delta_{max}=24.5$~GHz (obtained from the maximum Overhauser field \cite{paget1977} via $\delta=g\mu_{B}B_{n}/2\hbar$ with g-factor $g=-0.41$ \cite{sladkov2010}). The thermal equilibrium properties of the nuclear spin bath are well approximated by considering $N$ non-interacting spins $I$ with gyromagnetic ratio $\gamma$. Then $p$ and its variance $\sigma_{p}^2$ are in the high temperature limit $\hbar \gamma B_{ext}/k_{\text{B}}T\ll1$ (our experimental conditions) $p=\hbar\gamma B_{ext}(I+1)/3k_\text{B}T\approx 0$ and $\sigma_{p}^2=(I+1)/3IN-p^2$ \cite{coish2004}.

Because nuclear spin dynamics is slow as compared to the electron's, light interacting with the system sees a snapshot of the Overhauser shift taken from a distribution $P(\delta)$. A measurement on an ensemble of these systems should account for averaging over $P(\delta)$. The CPT lineshape of Fig.~\ref{fig:lambda}(b) arises from the transmittance, with a susceptibility that is averaged over $P(\delta)$,
\begin{equation}
T(\omega_{i})=\mathrm{exp}\left(-\rho\frac{\omega_{i}d}{c}\int_{-\infty}^{+\infty}P(\delta)\chi_{i}''(\omega_{i},\delta)\mathrm{d}\delta\right),
\end{equation}
where $d$ is the thickness of the medium, $\rho$ the density of donors, $c$ the speed of light, $i=1,2$ labels the laser fields. Here $\chi_{i}$ is the susceptibility for the laser field for a fixed $\delta$. It can be calculated from the Lindblad equation and depends on other system parameters implicitly \cite{onur2014}.
At thermal equilibrium $P(\delta)$ is a Gaussian centered at zero with variance $\sigma_{\delta}^2=\delta_{\text{max}}\sigma_{p}^2$. For $I=3/2$ and $N=10^{5}$ it has a width (FWHM) of
$2\sqrt{2 \, \mathrm{log}(2)}\sigma_{\delta}=136$~MHz, which roughly corresponds to the width of the measured CPT.

However, $P(\delta)$ can undergo changes when the electron spin is brought out of thermal equilibrium by optical orientation. An optically-induced electron spin polarization will in turn induce nuclear spin polarization via a hyperfine-mediated cross-relaxation process known as DNP. In Ref.~\cite{onur2014} it was described how the interplay between the laser-induced electron spin polarization near CPT resonance and DNP can change the shape of $P(\delta)$ by autonomous feedback control, leading to the formation of stable states for the nuclear spin polarization and offering the potential of reducing the variance $\sigma_{\delta}^2$. The essence of this method is pictured schematically in Fig.~\ref{fig:lambda}(c). It shows two distinct control regimes (color coded, red and blue) where both lasers are either red ($\Delta<0$) or blue ($\Delta>0$) detuned from the excited state. The change in laser coupling strength with $\delta$ is asymmetric when $\Delta\neq0$ (one laser approaches resonance while the other moves away from it). For a single system with a particular Overhauser shift this causes a sharp change in the optically-induced electron spin polarization $\braket{S_{z}}-\braket{\overline{S}_{z}}$ (where the overbar implies that the expectation value is taken at thermal equilibrium), shown in the middle panels as a function of $\delta$ (the Overhauser shift is here normalized to $\Gamma_{3}$). The blue and red dots indicate stable points, where $\braket{S_{z}}=\braket{\overline{S}_{z}}$ and $\partial/\partial\delta (\braket{S_{z}}-\braket{\overline{S}_{z}})<0$. We thus expect $P(\delta)$ to evolve from the initial Gaussian to either a distribution with two maxima, or to a distribution with one maximum. Such steady-state distributions are non-thermal and can thus have reduced fluctuations if the system's feedback response (slope of $\braket{S_{z}}-\braket{\overline{S}_{z}}$ near the stable point) is strong enough \cite{onur2014}.

\begin{figure}
\centering
\includegraphics[width=7cm]{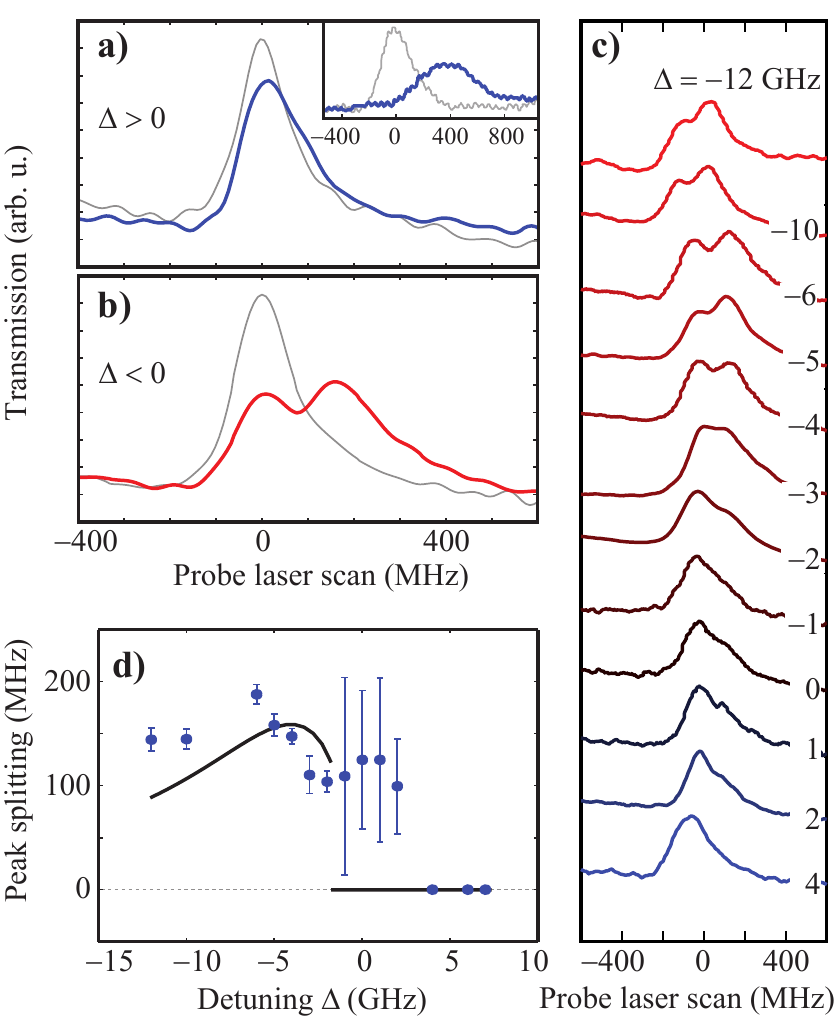}
\caption{CPT signatures of DNP feedback. (a) The CPT peak in the transmission signal as a function of probe laser frequency, before (gray) and after (blue) a DNP pumping period with two lasers fixed on two-photon resonance (Fig.~1(c)) and detuning $\Delta=+4$~GHz. The inset shows how DNP pumping by one laser (on the $\ket{\uparrow}-\ket{D^{0}X}$ transition) causes a shift of the CPT resonance. Two-laser pumping stabilizes the nuclear spin polarization at its thermal equilibrium value (here without observing a significant narrowing).
(b) Results obtained as for panel~(a), but with $\Delta=-6$~GHz. In this case the CPT peak after DNP pumping (red) shows a splitting.
(c) CPT traces taken after DNP pumping, for various values of $\Delta$.
(d) Values of the peak splitting, obtained from traces as in panel~(c). All data was taken with both laser intensities stabilized at values of about 3~${\rm W cm^{-2}}$  (for DNP pumping and CPT probing). Black line: simulation with parameters as in Ref.~\cite{onur2014}, except $\Gamma_{d}/\overline{\Gamma}_{h}=4000$ and $\gamma_{3}=20$~GHz.}
\label{fig:2}
\end{figure}

We investigate this interplay between CPT and DNP for the donor-bound electrons in GaAs by monitoring the changes in the CPT lineshape induced by two-laser optical pumping, with both lasers at equal intensity near two-photon resonance. Figures~\ref{fig:2}(a,b) show the CPT lineshape before (gray lines) and after 10~min of optical pumping with blue- and red-detuned lasers.
While scanning over the ensemble CPT peak, the probe laser meets exact two-photon resonances (near-ideal CPT peaks) of individual electrons for a range of $\delta$-values. The susceptibility is thus proportional to the number of electron spins experiencing a particular Overhauser shift $\delta$, hence reflecting the underlying nuclear spin distribution.
The nuclear spin distribution stabilizes as predicted in both cases, observed as a non-shifted single CPT peak in Fig.~\ref{fig:2}(a) and a non-shifted split CPT peak in Fig.~\ref{fig:2}(b) (the splitting directly reflects the doubly peaked $P(\delta)$ of Fig.~\ref{fig:lambda}(c)). This is in clear contrast with a CPT peak recorded after 10~min of single-laser optical pumping (inset Fig.~\ref{fig:2}(a)), which shifts the CPT peak by $\sim$400~MHz since DNP gives here a net nuclear spin polarization.

The lineshape in the main panel of Fig.~\ref{fig:2}(a) remains similar, while a narrower and higher CPT peak is expected if the width of the stabilized $P(\delta)$ would indeed be reduced. In Ref.~\cite{onur2014} it was pointed out that for an open system the narrowing by the feedback mechanism is in competition with nuclear spin diffusion. For donors in GaAs this plays a stronger role than for quantum dots, where a material barrier surrounding the dot suppresses this spin diffusion. Not observing a narrowing of the CPT peak is also due to non-uniform laser intensities for the electron ensemble (further discussed below).

Figure~\ref{fig:2}(c) shows the transition from red- to blue-detuned two-laser pumping, for a range of detunings $\Delta$.  Splittings in these CPT peaks are analyzed in Fig.~\ref{fig:2}(d), obtained by fitting two Gaussians to each CPT peak. Where the fit does not improve with respect to a single-Gaussian fit we take the splitting to be zero. The data reproduces the essential features of the model \cite{onur2014} (black line), showing a discontinuous transition and a maximum splitting when the pump lasers are tuned to slope of the transition line at $\Delta \approx -5$~GHz, where the response to a shift of $\delta$ is largest. We analyzed that this transition is a unique feature that confirms the dominance of the electron spin for the relevant DNP mechanism \cite{onur2014}. For $\Delta \gtrsim 0$ there is no good match, but the fitting also yields larger error bars. We attribute this to inhomogeneous broadening in the optical transitions (effective spread in detunings $\Delta$) which prevents all systems from making the transition simultaneously.

\begin{figure}
\centering
\includegraphics[width=7cm]{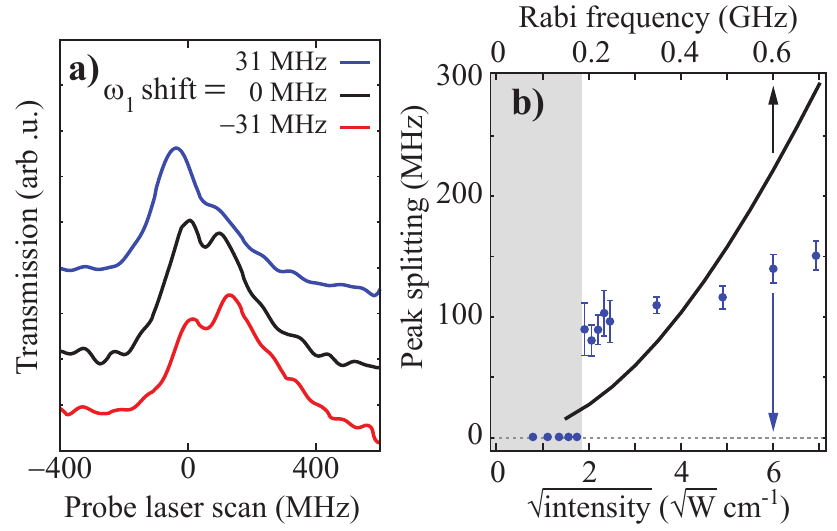}
\caption{(a) CPT traces after DNP pumping with two lasers at $\Delta \approx -3.5$~GHz. The trace labeled $\omega_1=0$~MHz is taken after pumping on exact two-photon resonance. The blue (red) trace is measured after pumping with $\omega_1=+(-)31$~MHz detuned from exact two-photon resonance (see Fig.~1(c)).
(b) The peak splitting in CPT traces after pumping with $\Delta \approx -4$~GHz and exact two-photon resonance, as a function of the intensity of the two lasers (keeping the intensity ratio fixed near 1). The gray background shows the range where the CPT peak shape was analyzed as a single peak. Significant double-peak character was observed for the total laser intensity above $\sim$3~${\rm W cm^{-2}}$. CPT traces were all taken with both lasers intensities at $\sim$3~${\rm W cm^{-2}}$. Black line: simulation with same parameters as in Fig.~\ref{fig:2}, the top axis shows the Rabi frequency corresponding to the simulation \cite{onur2014}.}
\label{fig:3}
\end{figure}

We now focus on the control regime $\Delta<0$ to examine the dependence of the stabilization on the control parameters during the optical pumping phase. Figure~\ref{fig:3}(a) shows the importance of carefully tuning the relative frequencies for getting a balanced distribution. A detuning as small as 31~MHz for one of the lasers gives a significant shift within $P(\delta)$ to either one of the stable states. Figure~\ref{fig:3}(b) shows values for the splitting as a function of the laser powers (varied simultaneously). The splitting shows a discontinuous onset and subsequent increase due to power broadening of the CPT peak. The data qualitatively matches the prediction (\cite{onur2014}, black line) but the slope is lower than the simulation. We attribute this to standing wave patterns in the GaAs layer (which acts as a weak cavity). The patterns for the two lasers do not fully overlap since they differ in frequency. This prohibits addressing the entire ensemble with equal laser intensities, and gives for the ensemble an averaged, less effective feedback mechanism. This also provides a limitation for the amount of CPT-peak narrowing in the blue-detuned case. The narrowing effect relies on carefully balanced laser intensities, and this is compromised due to the intensity variation inside the sample. Studying the achievable narrowing of $P(\delta)$ requires an experiment with uniform intensities for the ensemble.


\begin{figure}
\centering
\includegraphics[width=7cm]{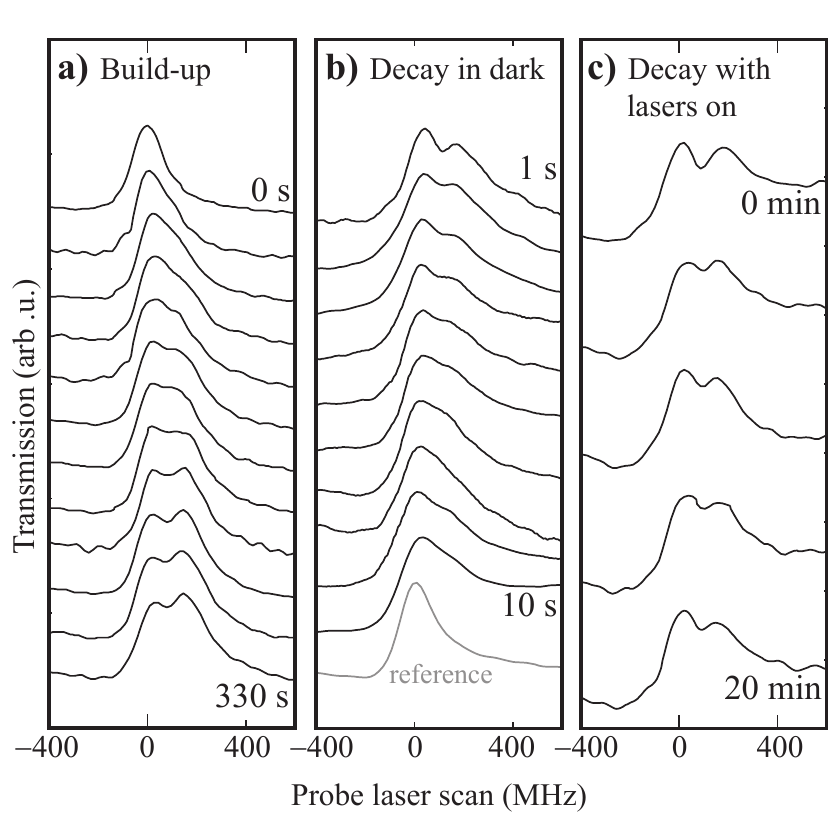}
\caption{Time evolution of build-up (a) and decay (b,c) of stabilized nuclear spin ensembles, measured during and after the DNP pumping period (subsequent traces top to bottom, as labeled). Panel~(b) gives a reference for the CPT peak before pumping.
The data in (a) is obtained from CPT scans of 1~s in between periods of 30~s DNP pumping with two lasers fixed at two-photon resonance ($\Delta=-4$~GHz).
Panel~(b) presents CPT scans of 1~s, taken after a fully dark period of 1~s (top trace) to 10~s (bottom trace) after DNP pumping ($\Delta=-4$~GHz).
The results in (c) are from continuously taking CPT scans of 1~s (only four traces shown), after a DNP pumping period at $\Delta=-2$~GHz.
All data was taken with both laser intensities stabilized at values of about 3~${\rm W cm^{-2}}$  (for DNP pumping and CPT probing).}
\label{fig:4}
\end{figure}

Figure~\ref{fig:4} presents time evolution of the effects. Figure~\ref{fig:4}(a) shows build-up of the splitting, obtained by taking CPT traces during the optical pumping phase every 30~s (each trace is collected within 1~s). The splitting stabilizes after approximately 4~min. Figure~\ref{fig:4}(b) shows decay of the splitting. It consists of traces collected after the optical pumping phase. After 10~min of optical pumping (repeated before each trace) the system is kept in the dark for a time ranging from 1 to 10~s. The splitting fades away in seconds, consistent with the relaxation of the lattice nuclear spins by spin diffusion away from the electron \cite{paget1982}. However, when CPT scans are taken continuously after the optical pumping phase the splitting decays much slower and persists up to at least 20~min (Fig.~\ref{fig:4}(c), we verified that taking such scans without the preceding pumping phase does not induce a splitting). We attribute this to a suppression of the spin diffusion while the system is illuminated: under optical excitation (during CPT scans) the electron spin is most of the time significantly polarized and this suppresses nuclear spin diffusion because it creates an inhomogeneous Knight field for the surrounding nuclear spins \cite{deng2005,gong2011}. This effect could be used to improve the strength of the feedback control and the amount of narrowing: if the temperature of the experiment would be lower or the magnetic field stronger (increased $\braket{\overline{S}_{z}}$) the thermal-equilibrium electron spin polarization can suppress nuclear spin diffusion.


Our results open the possibility to use the interplay between CPT and DNP to operate a mesoscopic spin system as a feedback loop that converges towards a well defined steady state, determined by laser power and detuning, with the possibility of reduced nuclear spin fluctuations and less electron spin dephasing. The mechanism is generally applicable to localized spins where DNP is dominated by electron-nuclear spin hyperfine coupling and can also be used for other paramagnetic defects, as ensembles or single systems. A notable example is the fluorine donor in ZnSe \cite{greilich2012,kim2014}, a II-VI material with dilute nuclear spins (in GaAs all atoms have non-zero nuclear spin). Nuclear spin diffusion, mediated by dipole-dipole interaction (inversely proportional to distance between nuclear spins to the power 6), will here be much less a limitation for narrowing.

\begin{acknowledgments}
We thank A.~U.~Chaubal and R.~S.~Lous for help and valuable discussions, and acknowledge financial support from the Dutch FOM and NWO, ERC Starting Grant 279931, the Research school
Ruhr-Universit\"{a}t Bochum, and the German programs
BMBF Q.com-H 16KIS0109, Mercur Pr-2013-0001, and
the DFH/UFA CDFA-05-06.

\end{acknowledgments}

\bibliographystyle{apsrev4-1}
\bibliography{splittingbibliography}

\end{document}